\documentclass[pdflatex,sn-mathphys-num]{sn-jnl}

\usepackage{graphicx}%
\usepackage[section]{placeins}

\usepackage{multirow}%
\usepackage{amsmath,amssymb,amsfonts}%
\usepackage{amsthm}%
\usepackage{mathrsfs}%
\usepackage[title]{appendix}%
\usepackage{xcolor}%
\usepackage{textcomp}%
\usepackage{manyfoot}%
\usepackage{booktabs}%
\usepackage{algorithm}%
\usepackage{algorithmicx}%
\usepackage{algpseudocode}%
\usepackage{listings}%

\theoremstyle{thmstyleone}%
\theoremstyle{thmstyletwo}%

\theoremstyle{thmstylethree}%

\usepackage{lineno}

\begin{document}

\title[Article Title]{Generative artificial intelligence improves projections of climate extremes}




\author[1,2,5]{\fnm{Ruian} \sur{Tie}}\email{ratie25@m.fudan.edu.cn}
\equalcont{These authors contributed equally to this work.}

\author[1,6]{\fnm{Xiaohui} \sur{Zhong}}\email{x7zhong@gmail.com}
\equalcont{These authors contributed equally to this work.}

\author[1,5]{\fnm{Zhengyu} \sur{Shi}}\email{zhengyushi@fudan.edu.cn}
\equalcont{These authors contributed equally to this work.}

\author*[1,2,5,6]{\fnm{Hao} \sur{Li}}\email{lihao\_lh@fudan.edu.cn}

\author[3,4]{\fnm{Bin} \sur{Chen}}\email{chen\_bin@fudan.edu.cn}

\author[1]{\fnm{Jun} \sur{Liu}}\email{23110240031@m.fudan.edu.cn}

\author*[4,2,5]{\fnm{Libo} \sur{Wu}}\email{wulibo@fudan.edu.cn}

\affil[1]{\orgdiv{Artificial Intelligence Innovation and Incubation Institute}, \orgname{Fudan University}, \orgaddress{\street{220 Handan Road}, \city{Shanghai}, \postcode{200433}, \state{Shanghai}, \country{China}}}
\affil[2]{\orgdiv{Shanghai Innovation Institute}, \orgaddress{\street{669-3 Huafa Road}, \city{Shanghai}, \postcode{200433}, \state{Shanghai}, \country{China}}}

\affil[3]{\orgdiv{Institute for Big Data}, \orgname{Fudan University}, \orgaddress{\street{220 Handan Road}, \city{Shanghai}, \postcode{200433}, \state{Shanghai}, \country{China}}}

\affil[4]{\orgdiv{MOE Laboratory for National Development and Intelligent Governance}, \orgname{Fudan University}, \orgaddress{\street{220 Handan Road}, \city{Shanghai}, \postcode{200433}, \state{Shanghai}, \country{China}}}

\affil[5]{\orgdiv{Shanghai Academy of AI for Science}, \orgaddress{\street{18-X3 Longyao Road}, \city{Shanghai}, \postcode{200433}, \state{Shanghai}, \country{China}}}

\affil[6]{\orgname{FuXi Intelligent Computing Technology Co., Ltd.}, \city{Shanghai}, \postcode{200030},  \country{China}}


\abstract{
Climate change is amplifying extreme events, posing escalating risks to biodiversity, human health, and food security.
Global climate models (GCMs) are essential for projecting future climate, yet their coarse resolution and high computational costs constrain their ability to represent extremes.
Here, we introduce FuXi-CMIPAlign, a generative deep learning framework for downscaling Coupled Model Intercomparison Project (CMIP) outputs.
The model integrates Flow Matching for generative modeling with domain adaptation via Maximum Mean Discrepancy loss to align feature distributions between training data (ERA5 reanalysis) and inference data (European Consortium-Earth), thereby mitigating input discrepancies and improving accuracy, stability, and generalization across emission scenarios. 
FuXi-CMIPAlign performs spatial, temporal, and multivariate downscaling, enabling more realistic simulation of compound extremes such as tropical cyclones (TCs).
Applied to the historical period (2005–2014), it reduces global 99th-percentile mean absolute errors by 26\%, 42\%, and 33\% for high temperature, extreme precipitation, and strong wind, respectively, and reproduces TC activity better aligned with ERA5.
Under future scenarios (2015–2100), FuXi-CMIPAlign projects pronounced increases in land area affected by high temperature and frequency of extreme precipitation under high-emission scenarios, along with up to 60\% rise in TC intensity and frequency over the Northwest and Northeast Pacific.
In contrast, strong wind events over land shows a counterintuitive weakening trend.
These results demonstrate that FuXi-CMIPAlign substantially improves CMIP6 projections of climate extremes, providing a robust generative framework for advancing climate risk assessment, mitigation and adaptation.
}

\maketitle

\section{Introduction}\label{sec1}

Climate change is amplifying extreme weather and climate events worldwide \cite{tripathy2023climate}.
Anthropogenic greenhouse gas emissions have disrupted the Earth’s climate system, driving more frequent and severe heatwaves \cite{raymond2020emergence}, cold spells \cite{ruosteenoja2025extreme}, heavy precipitation \cite{fowler2024precipitation}, agricultural droughts \cite{pizzorni2024droughts}, and tropical cyclones (TCs) \cite{calvin2023ipcc}.
Between 2016 and 2024, daily land temperature records show that extreme heat events occurred over four times more often than expected, while cold records declined by half \cite{fischer2025record}.
These unprecedented shifts threaten human health \cite{romanello20232023, grant2025global}, infrastructure \cite{adshead2024climate, liu2023global}, food security \cite{singh2023climate}, biodiversity \cite{kotz2025large}, and global economies \cite{sun2024global,kotz2022effect}.
Therefore, reliable climate projections are essential for effective mitigation and adaptation strategies \cite{newman2023global,el2024europe,xu2025quantifying}.

The Coupled Model Intercomparison Project (CMIP) \cite{meehl2000coupled} provides a foundation for global climate projections.
Since its launch in 1995, CMIP has coordinated systematic evaluation of coupled general circulation models (GCMs).
CMIP5 introduced Representative Concentration Pathways (RCPs), while CMIP6 extended this framework by incorporating Shared Socioeconomic Pathways (SSPs) through ScenarioMIP, enabling consistent simulations of emissions and socioeconomic trajectories to 2100 and facilitating integrated assessment of climate risks \cite{o2016scenario}.
These advances have greatly enhanced the scientific and policy relevance of climate projections.

However, widely used GCMs, such as the European Consortium-Earth (EC-Earth) \cite{doscher2021ec} remain constrained by coarse spatial resolution (typically $> 100$ km), and imperfect parameterizations, constrained in persistent systematic biases.
Although the CMIP6 sub-project HighresMIP provides higher-resolution simulations, it covers only limited emission scenarios and still underestimates the frequency and intensity of extremes \cite{haarsma2016high}.
Despite progress from CMIP5 to CMIP6, improvements in reproducing historical climate extremes remain modest, and key biases persist.
For example, CMIP6 reduces warm bias over mid-latitudes in South America and Asia \cite{kim2020evaluation}, yet strong cold bias remain at high latitudes, along with wet biases over Tibet, northwest China, and the Maritime Continent, and dry biases in regions of frequent rainfall such as southern China \cite{dong2021evaluation}.
Moreover, the coarse spatiotemporal resolution of most GCMs also limits their ability to represent local-scale extremes such as TCs \cite{roberts2020projected}, which depend sensitively on regional ocean–atmosphere interactions.
Thus, Low-resolution models struggle to reproduce these regional climate patterns and their impacts on TC tracks and intensities \cite{sainsbury2022can}.

Because extreme events typically emerge at local scales, their reliable simulation requires medium- to fine-scale representation.
To bridge this gap, downscaling methods are widely applied to enhance GCM resolution and improve extreme event projections.
Downscaling transforms coarse GCM outputs into high-resolution local fields through two main approaches: dynamical and statistical downscaling \cite{hewitson1996climate}.
Dynamical downscaling employs high-resolution regional climate models (RCMs) nested within GCMs, but these simulations inherit biases from the driving models, depend on boundary conditions, and demand substantial computational resources, limiting global application \cite{xu2019dynamical}.
Statistical downscaling instead establishes empirical relationships between historical observations and GCM outputs \cite{wilby2013statistical}.
Classical approaches such as quantile mapping (QM) \cite{maraun2013bias} and delta methods \cite{hay2000comparison} remain popular, but assume linear bias evolution and poorly capture nonlinear climate responses, often underestimating extremes and require large ensembles for robust results \cite{tefera2024evaluation,miller2025statistical}.
Both RCM-based and traditional statistical approaches are computationally intensive and inefficient for global-scale extreme event projections.


Recent advances in reanalysis datasets, notably the fifth generation reanalysis (ERA5) from the European Centre for Medium-Range Weather Forecasts (ECMWF) \cite{hersbach2020era5}, together increasing computing power, have positioned deep learning models as the new state-of-the-art in statistical downscaling \cite{sun2024deep}.
A key methodological shift is the emergence of transferable models that can be trained on one dataset and directly applied to others, greatly accelerating research and broadening applicability across GCMs.
Early efforts, such as Deep Statistical Downscaling (DeepSD), trained models on coarse inputs derived from high-resolution reference data but applied them directly to raw GCM outputs, leading to mismatches between training and inference conditions \cite{vandal2017deepsd}.
To mitigate this, Bano et al. proposed Deep Ensemble Statistical Downscaling (DeepESD), which reduced systematic biases across GCMs through conventional bias correction before generating ensemble downscaled results \cite{bano2022downscaling}.
Although this improved generalization, it assumed that historical bias corrections remained valid under future scenarios.
More recently, Hess et al. introduced a Consistency Model (CM) that injected noise to degrade small-scale features, and align training and inference in the frequency domain \cite{hess2025fast}.
However, frequency domain adaptation alone cannot fully capture the joint statistical properties of climate data.
These developments mark a transition from deterministic, bias-corrected ensemble approaches toward generative modeling with zero-shot capabilities.
A key challenge in this transition is domain adaptation, ensuring consistent model performance when training and inference data differ.
Yet the role of domain adaptation in simulating extremes remains underexplored.
Moreover, most studies have focused on single-variable downscaling (e.g., temperature or precipitation), whereas compound extremes such as TCs depend on multiple interacting variables (e.g., wind speed, mean sea level pressure, and geopotential) and require sub-daily temporal resolution.
For instance, TC tracks are more reliably identified at six-hourly intervals than at daily scales (Supplementary Fig. 1).

Here, we present FuXi-CMIPAlign, a generative deep learning downscaling framework for CMIP that integrates Flow Matching with domain adaptation to mitigate input discrepancies and enhance accuracy, stability, and generalization across emission scenarios.
FuXi-CMIPAlign conducts spatial, temporal, and multivariate downscaling, transforming global daily fields at 0.70$^\circ$ resolution into 6-hourly fields at 0.25$^\circ$, and simultaneously downscaling five land and two atmospheric variables to capture high temperatures, extreme precipitation, strong winds, and compound extremes such as TCs.
Through domain adaptation using Maximum Mean Discrepancy (MMD) loss, FuXi-CMIPAlign aligns feature distributions between training (ERA5) and inference (EC-Earth) domains, thereby improving cross-domain consistency.
When applied to the historical period (2005-2014), FuXi-CMIPAlign substantially outperforms EC-Earth and noise-based alignment methods in reproducing historical extremes.
Under future scenarios (2015–2100), FuXi-CMIPAlign projects intensifying risks, showing increasing land area affected high temperature, more frequent extreme precipitation, and denser TC tracks over the Northwest and Northeast Pacific.
These results underscore the escalating impacts of climate change and the importance of robust, generative downscaling frameworks for informing adaptation strategies and policy decisions.


\section{Results}
\label{sec2}

In this section, we present results from two perspectives.
First, we evaluate model performance during the historical period (2005–2014) by comparing 99th-percentile mean absolute error (MAE) using ERA5 as the reference.
Second, we assess future extremes under different emission scenarios (2015–2100), focusing on the spatiotemporal evolution of area fraction and regional mean frequency of high temperatures, extreme precipitation, and strong wind events.
These extremes are defined using 2-meter temperature (T2M), 6-hour accumulated total precipitation (TP), and 10-meter wind speed (WS10M).
We also analyze the distribution and intensity of TCs.
In all experiments, EC-Earth from CMIP6 serves as the baseline model and provides input to FuXi-CMIPAlign.
The term noalign denotes experiments without domain adaptation, whereas FuXi-CMIPAlign refers to the proposed domain-adapted framework.

\subsection{Performance during the historical period}
\label{subsec1}

\begin{figure}[h]
\centering
\includegraphics[width=1\textwidth]{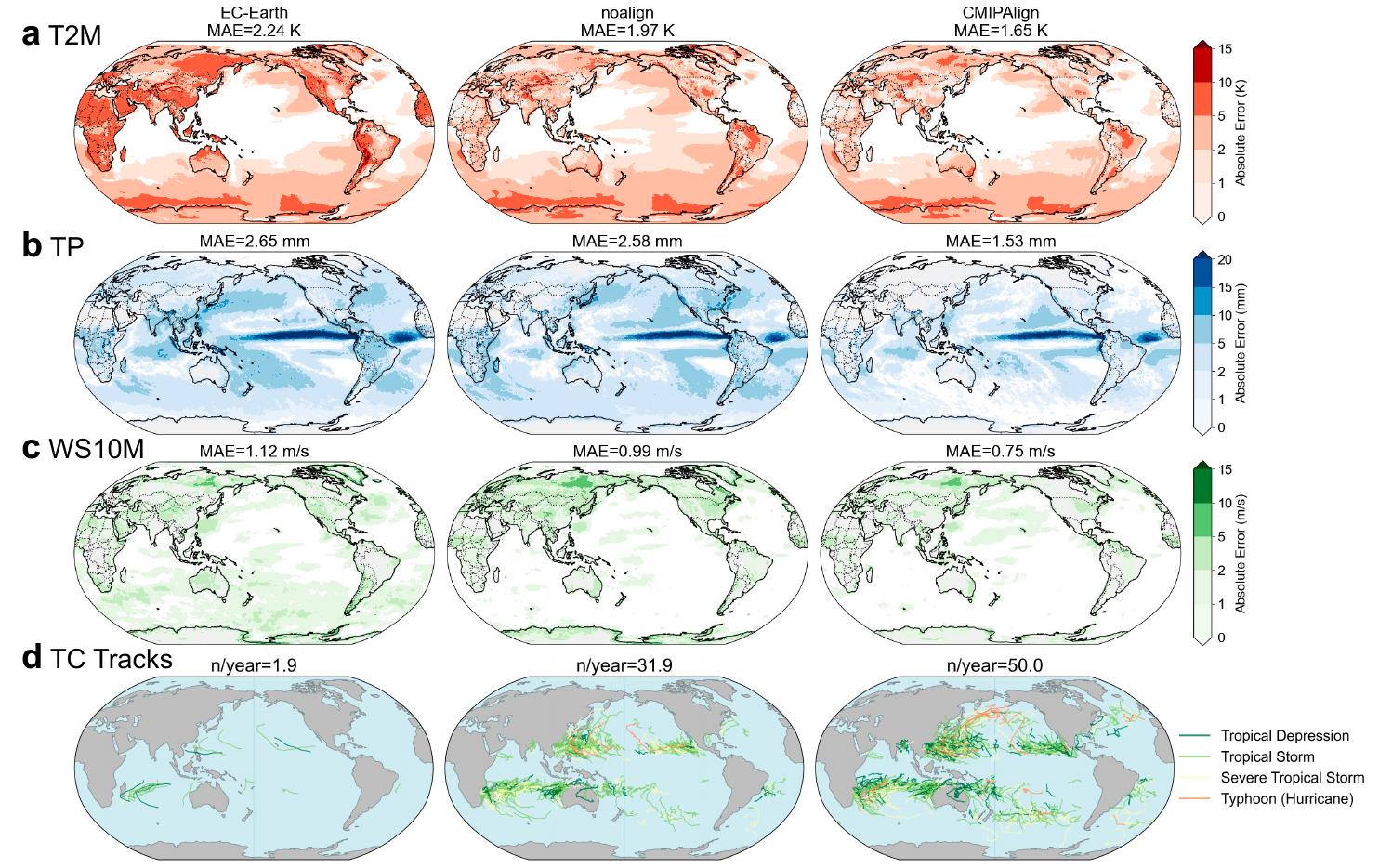}
\caption{Spatial distribution of 99th-percentile mean absolute error (MAE) and accumulative tropical cyclone (TC) tracks from 2005 to 2014. \textbf{a}-\textbf{c,} show MAE for 2-meter temperature (T2M), 6-hour accumulated total precipitation (TP), and 10-meter wind speed (WS10M), respectively. \textbf{d} shows TCs, with line colors indicating intensity categories and numbers (n/year) denoting the annual mean number of TC tracks detected over the 10-year period. Columns correspond to European Consortium-Earth (EC-Earth), noalign, and FuXi-CMIPAlign, from left to right.}\label{FIG1}
\end{figure}

Fig. \ref{FIG1} presents the spatial distribution of the 99th-percentile mean absolute error (MAE) for three key variables—T2M, TP, and WS10M—as well as the accumulative TC tracks during the historical period from 2005 to 2014.
The figure compares the performance of three models: EC-Earth, the noalign method, and the proposed FuXi-CMIPAlign. The noalign excludes domain adaptation, applying the model trained by ERA5 directly to EC-Earth data.
The results in Fig. \ref{FIG1} show that FuXi-CMIPAlign consistently outperforms the other methods in capturing extremes and reducing errors, with particular strong improvements in TC detection and the simulation of extreme events.

For extreme heat, FuXi-CMIPAlign achieves an MAE of 1.65 K, outperforming EC-Earth (2.24 K) by 26\% and noalign (1.97 K).
Spatially, it reduces errors in regions prone to heat extremes, such as tropical and subtropical areas across Africa, northern South America, and Southeast Asia.
EC-Earth exhibits large biases in these areas, typically underestimating extremes over land and overestimating them over ocean (see 99th-percentile biases in Supplementary Information Figure 7a).
Although noalign shows some improvement, it fails to capture heat extremes effectively.
In contrast, FuXi-CMIPAlign more accurately reproduces their spatial distribution, demonstrating the role of domain adaptation in reducing biases and improving model performance.

For extreme precipitation, FuXi-CMIPAlign yields an MAE of 1.53 mm, significantly lower than 2.65 mm for EC-Earth (a 42\% improvement) and 2.58 mm for noalign.

Globally, it better captures local precipitation patterns, especially in the tropics.
However, in the Intertropical Convergence Zone (ITCZ), all models show systematic negative biases.
EC-Earth and noalign show stronger underestimating, especially in the central Pacific Ocean and the Atlantic Ocean, whereas FuXi-CMIPAlign more faithfully represents regional precipitation (see Supplementary Information Figure 7b).

Regarding strong winds, FuXi-CMIPAlign reduces MAE to 0.75 m/s, representing a 33\% improvement compared with 1.12 m/s for EC-Earth and 0.99 m/s for noalign.
Improvments are most pronounced in coastal and oceanic regions, particularly in TC-prone basins such as the Northwest Pacific, North Atlantic, and Indian Ocean.
The wind speed simulation errors in these regions are significantly lower compared to the other two models.
EC-Earth exhibits substantial deviations along TC tracks, while noalign reduces but does not eliminate errors over land surface.
FuXi-CMIPAlign captures the spatial distribution of strong winds more accurately, particularly in high-intensity events.

In TC detection, FuXi-CMIPAlign also demonstrates clear advantages in simulating extremes that rely on multiple variables.
On average, it detects 50 tracks per year, closely matching ERA5 (62 tracks per year, refer to Supplementary Information Figure 1, left), compared with only approximately 2 for EC-Earth and 32 for noalign.
FuXi-CMIPAlign also recovers the full spectrum of TC intensities (see Table \ref{TC_intensity_scale}), from tropical depressions to typhoons, comprehensively reflecting cyclone variability.
In contrast, EC-Earth performs poorly in both frequency and intensity classification, while noalign improves detection but underestimates strong TCs.
Overall, FuXi-CMIPAlign offers more comprehensive simulations of both track distribution and intensity evolution.
A limitation shared by all models is that the detected TC tracks exhibit discontinuities or abrupt changes, appearing less smooth than those in ERA5.
This issue is likely due to the use of daily training data, which preserves temporal consistency within a single day (e.g., 00, 06, 12, and 18 UTC), but introduces inconsistencies between consecutive days (e.g., 18 UTC on day t and 00 UTC on day t+1).
As a result, some TC tracks show unrealistic jumps that exceed the maximum feasible displacement within a 6-hour interval, resulting in fragmented TC trajectories and potential under-detection of TCs.

We further conducted ablation experiments to analyze model performance and validate the proposed framework (Supplementary Information Section 3).
First, we compared univariate with multivariate modeling for extremes and TC detection.
Multivariate modeling substantially outperforms univariate modeling, lowering 99th-percentile MAE (Supplementary Information Table 2) and increasing the number of detected TC tracks (Supplementary Information Figure 2).
Second, we tested end-to-end training, which directly conditions EC-Earth data with ERA5 as the diffusion denoising target.
While EC-Earth and ERA5 share the same timestamps, their spatial patterns differ significantly, producing poor spatial consistency when EC-Earth data are used as the condition and ERA5 as the target (Supplementary Information Figure 9).
This results in significant degradation of the 99th-percentile MAE (Supplementary Information Table 4).
Furthermore, as elaborated in Supplementary Information Section 3.2, we compared FuXi-CMIPAlign with the noise-based aligned method proposed by Hess et al. \cite{hess2025fast}.
Across energy spectrum analysis, percentile MAE, and TC detection, FuXi-CMIPAlign achieves closer agreement with ERA5.
It best reproduces the energy spectrum below 512 km for T2M, U10M, V10M, MSL, and TP (Supplementary Information Figure 6).
Compared to the noise-based method, FuXi-CMIPAlign reduces global 99th-percentile MAE values by 14\% (T2M), 40\% (U10M), and 13\% (MSL) (Supplementary Information Table 3), and increases the global annual average number of detected TCs by 38\% (Supplementary Information Figure 8).

In summary, FuXi-CMIPAlign demonstrates remarkable advantages in simulating extremes during the historical period.
Across T2M, TP, WS10M, and TC Tracks, it reduces errors while accurately capturing extreme characteristics.
Its strength is most pronounced in TC detection and intensity classification, underscoring its potential for high-resolution climate analysis and climate change impact assessment.
Importantly, because the carbon emission trends and background conditions during training differ significantly from those in the testing period, the model's strong performance highlights its generalization ability across emission scenarios.
FuXi-CMIPAlign provides a powerful tool for advancing the understanding of extreme weather mechanisms and projecting their evolution under future climate change.

\FloatBarrier

\subsection{Projected land Extremes across emission scenarios}\label{subsec2}



\begin{figure}[h]
\centering
\includegraphics[width=1\textwidth]{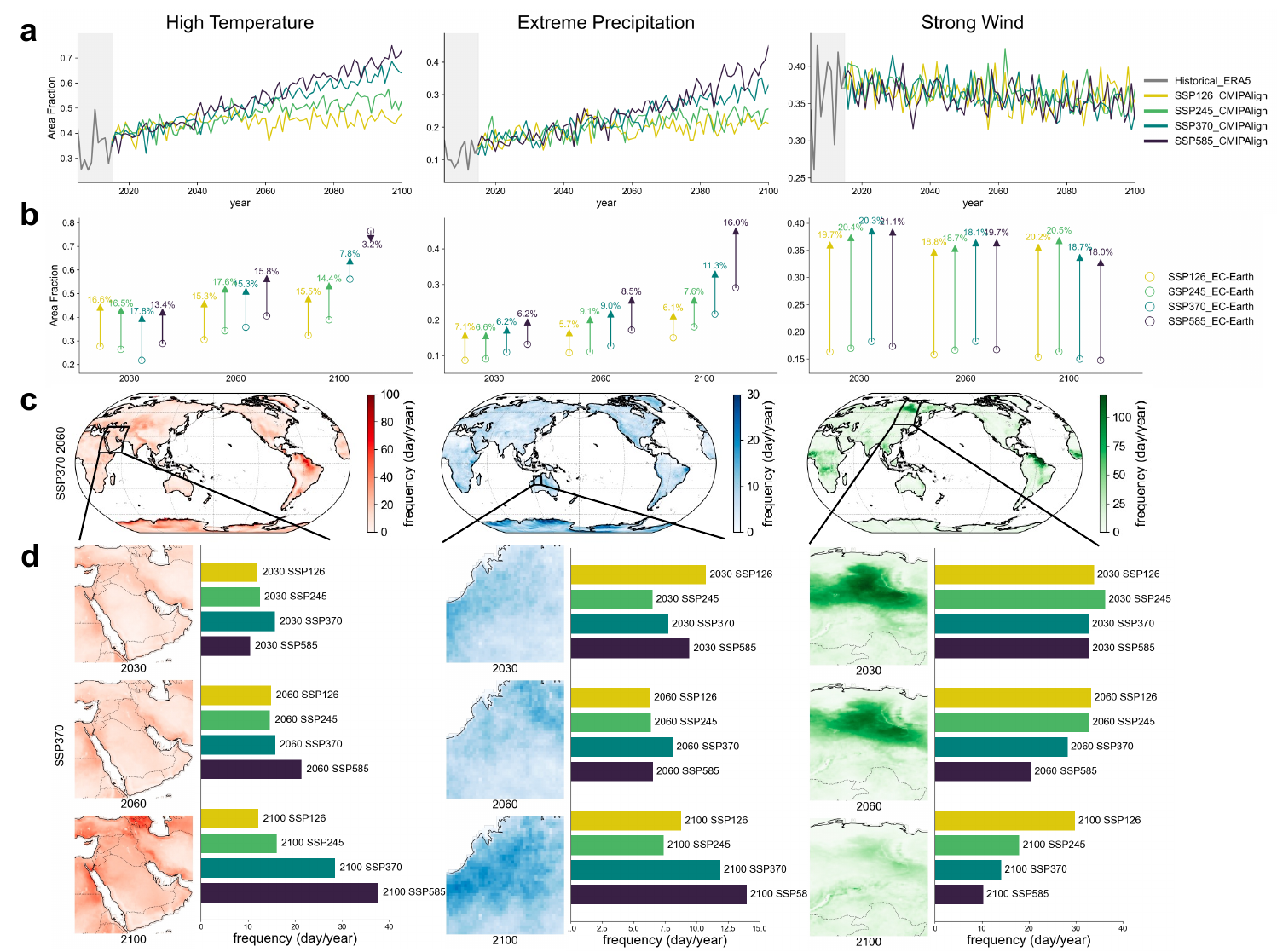}
\caption{Spatiotemporal trends of high temperature, extreme precipitation, and strong winds over land under emission scenarios (2015–2100). These events are defined by the 99th percentile thresholds of 2-meter temperature (T2M), 6-hour accumulated total precipitation (TP), and 10-meter wind speed (WS10M), respectively. \textbf{a,} Temporal evolution of land area fraction affected by each event type. The gray line represents historical observations based on ERA5 data (2005–2014), while the colored lines indicate projections from FuXi-CMIPAlign across SSP126 to SSP585 scenarios (2015–2100). \textbf{b,} Correction relative to European Consortium-Earth (EC-Earth): circles denote EC-Earth, arrows point to FuXi-CMIPAlign, and arrow length indicates the magnitude of correction. Values above arrowheads quantify corrections, with positive values indicating increases and negative values decreases. \textbf{c,} Global land frequency of event distribution in 2060 under SSP370. \textbf{d,} Spatial frequency of extreme events in three climate-sensitive regions (Middle East, Northwestern Australia, and Siberia) under SSP370, along with annual frequencies of extreme events for 2030, 2060, and 2100 under different scenarios.}\label{FIG2}
\end{figure}

Fig. \ref{FIG2}a shows the temporal evolution of land area fractions (refer to section \ref{subsec7} for how it is calculated) affected by these extreme events over time.
The gray line represents historical observations from ERA5 (2005–2014), while the colored lines show FuXi-CMIPAlign projections under four emission scenarios (SSP126 to SSP585) for 2015–2100.
The affected land area increases across all scenarios for both high temperature and extreme precipitation, with the most pronounced growth under the high-emission scenario (SSP585).
This indicates a substantial expansion of extreme heat and precipitation coverage in a warming climate, reflecting intensified warming and disruption of the global water cycle.
In contrast, the area fraction of strong winds over land shows a slight decline over time, especially under high-emission pathways such as SSP585.


Fig. \ref{FIG2}b highlights the land area fractions correction of FuXi-CMIPAlign relative to EC-Earth, where circles denote EC-Earth, arrows point
to FuXi-CMIPAlign, and arrow length indicates the magnitude of correction. Values above arrowheads
quantify corrections, with positive values indicating increases and negative values decreases. It reveals that EC-Earth systematically underestimates the area fraction of all extremes across different scenarios, although the magnitude varies.
Relative to EC-Earth, FuXi-CMIPAlign detects up to 17.8\% more high-temperature events (2030, SSP370), 16.0\% more extreme precipitation (2100, SSP585), and 20.5\% more strong winds (2100, SSP245).
For strong winds, FuXi-CMIPAlign achieves improvements exceeding 18\% across all periods and scenarios, demonstrating consistent skill improvements.

Fig. \ref{FIG2}c presents the spatial distribution of global extreme events under SSP370 in 2060. The spatial distribution of global extreme event frequencies varies substantially across regions.
In 2060 under the SSP370, a pathway most consistent with the current socio-economic and environmental trends, high temperature extremes are primarily concentrated in East Asia, northeastern South America, and Antarctica. Extreme precipitation events are mainly distributed in Antarctica, eastern South America, and Australia, and strong winds are predominantly concentrated in central Africa, Siberia, and northern South America. 

Based in Fig. \ref{FIG2}c, Fig. \ref{FIG2}d further examines three climate-vulnerable regions, such as Middle East, northwestern Australia, and Siberia, and presents the frequency distribution and annual occurrence of extreme events under different emission scenarios for 2030, 2060, and 2100.
The Middle East demonstrates a pronounced rise in extreme temperature frequency, underscoring its high sensitivity to warming.
Northwestern Australia exhibits heterogeneous changes in extreme precipitation, with rapid increases in some areas and slower shifts elsewhere.
Siberia displays a clear decline in strong wind frequency and coverage over time, especially under high-emission scenarios (SSP585).
The accompanying bar charts confirm that high temperature and extreme precipitation intensify under high-emission scenarios (e.g., SSP585) compared to low-emission scenarios (e.g., SSP126), whereas the frequency of strong wind events decrease consistently across all scenarios.

Finally, Supplementary Figures 10–15 extend the spatial distribution analysis of extreme-event frequencies for 2030, 2060, and 2100 across all emission scenarios using FuXi-CMIPAlign.

\FloatBarrier

\subsection{Detectable tropical cyclones in future emission scenarios}\label{subsec3}

\begin{figure}[h]
\centering
\includegraphics[width=1\textwidth]{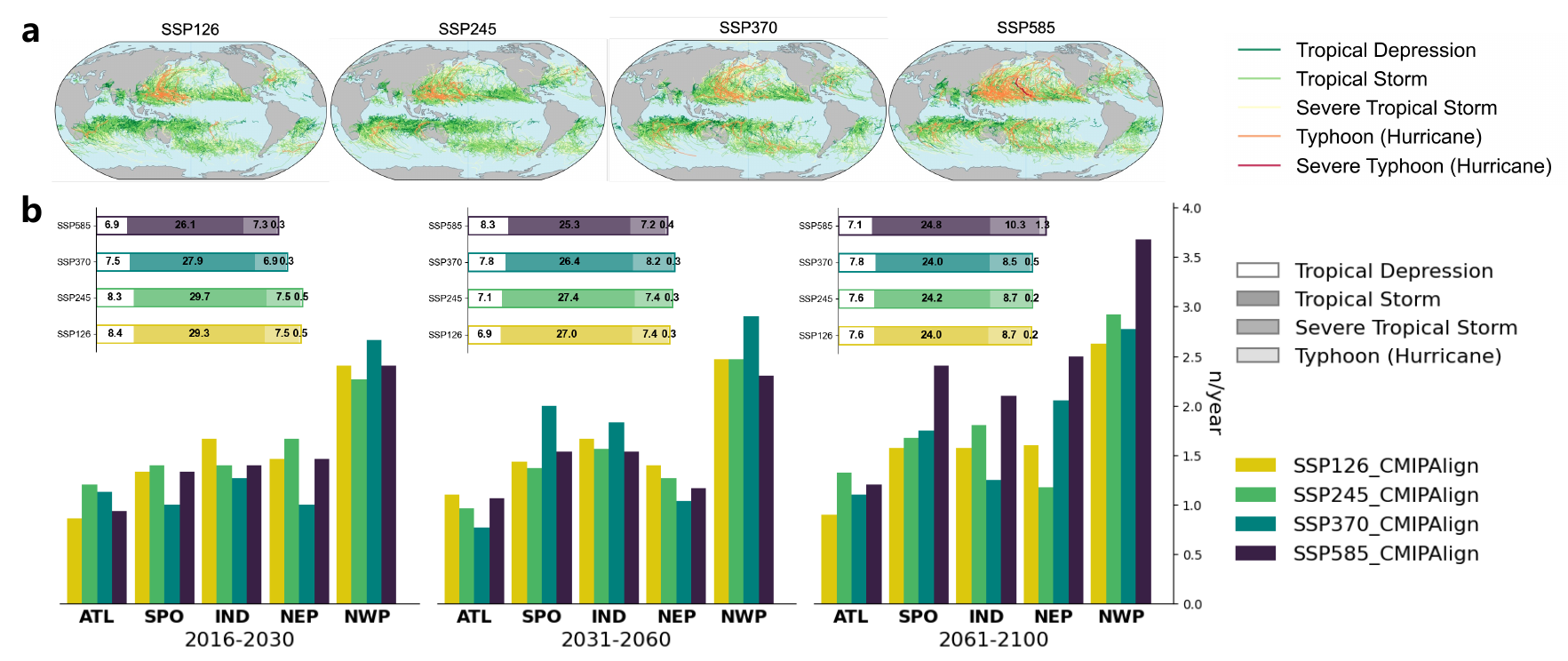}
\caption{Detectable tropical cyclone (TCs) tracks under emission Scenarios (2015-2100). \textbf{a,} Cumulative spatial distribution of TC tracks detected with FuXi-CMIPAlign under four emission scenarios (SSP126, SSP245, SSP370, SSP585) from 2015 to 2100. The color of TC tracks transitions from dark green to dark red, representing the intensity from tropical depression to severe typhoon (hurricane). \textbf{b,} The upper stacked graph shows the incremental annual mean number of TC tracks detected by FuXi-CMIPAlign relative to European Consortium-Earth (EC-Earth) across four scenarios, with colors indicating TC intensity categories. Bars in different colors represent different SSP scenarios, while different shades of the same color represent different intensities of TCs. The lower bar chart summarizes the annual mean number of Severe Tropical Storms and Typhoons (or Hurricanes) simulated in major ocean basins during three periods (2016–2030, 2031–2060, 2061–2100), with frequency statistics shown for regions of frequent TC activity. Basin abbreviations: ATL (Atlantic Ocean), SPO (South Pacific Ocean), IND (Indian Ocean), NEP (Northeast Pacific Ocean), NWP (Northwest Pacific Ocean).}\label{FIG3}
\end{figure}

TCs are fundamentally driven by oceanic heat release and exert destructive impacts through extreme precipitation and strong winds, making them closely associated with several key meteorological extremes.
FuXi-CMIPAlign demonstrates higher precision in capturing high temperature, extreme precipitation, and strong winds, thereby offering a substantial advantage in detecting TCs. Fig. \ref{FIG3} illustrates the spatial distribution of detectable TC tracks under different emission scenarios from 2015 to 2100.

Figure \ref{FIG3}a presents the cumulative spatial distribution of TC tracks detected by ClimAlign across four emission scenarios (SSP126, SSP245, SSP370, SSP585).
The tracks are color-coded from light to dark to represent increasing intensity, ranging from tropical depression to tropical storm, severe tropical storm, typhoon (or hurricane), and severe typhoon (or severe hurricane).
Under different emission scenarios, the tracks and intensity distribution exhibit notable regional differences.
For instance, under SSP126 (low-emission), TC tracks distributions remain relatively stable with limited spatial variation compared to historical baseline.
In contrast, under SSP585 (high-emission), both the density and intensity of TC tracks increase significantly, particularly in the Northwest and Northeast Pacific. 

The upper stacked bar chart in Fig. \ref{FIG3}b shows the annual average increase in TC tracks detected by FuXi-CMIPAlign compared to EC-Earth under the four emission scenarios, with colors indicating different TC intensity categories. Compared with EC-Earth, FuXi-CMIPAlign consistently detects more TC tracks across all emission scenarios, with notable improvements for Severe Tropical Storms and Typhoons (or Hurricanes).
Across the periods of 2016–2030, 2031–2060, and 2061–2100, FuXi-CMIPAlign identifies on average 43.7, 41.9, and 41.4 more tracks per year, respectively, than EC-Earth.
For Tropical Storms, the increase is nearly 25 times relative to EC-Earth, while for Typhoons (or Hurricanes), undetectable by EC-Earth (Supplementary Information Figure 16), FuXi-CMIPAlign detects about 0.5 tracks per year.
These results highlight the model's enhanced ability in downscaling TC-related variables, enabling a more accurate representation of TC activity under climate change.

The lower bar chart in Fig. \ref{FIG3}b summarizes the annual average number of severe tropical storms and typhoons (or hurricanes) across major ocean basins during three distinct periods (2016–2030, 2031–2060, 2061–2100). The frequency statistics focus on regions with high TC activity, with ocean basin abbreviations including ATL (Atlantic), SPO (South Pacific), IND (Indian Ocean), NEP (Northeast Pacific), and NWP (Northwest Pacific). For Severe Tropical Storms and Typhoons (or Hurricanes), which pose significant socio-economic risks, TC frequency exhibits a clear increasing trend under stronger forcing.
Under the high-emission scenario (SSP585), the number of TCs increases substantially in the Northeast Pacific (2.4 tracks per year) and Northwest Pacific (3.7 tracks per year) during 2061–2100 corresponding to a 60\% and 48\% increases relative to 2016–2030.
In contrast, changes in the Atlantic and Indian Ocean are relatively minor.
This regional disparity is likely related to the varying impacts of climate change on regional oceanic and atmospheric conditions.
For instance, stronger sea surface temperature increases under the high-emission scenario (Supplementary Information Figure 13), as warmer oceans supply more energy for TC genesis and intensification.
These results underscore the need for strengthened disaster risk management in vulnerable coastal regions, especially under high-emission scenarios, where TC activity is projected to intensify and pose greater threats to ecosystems and socio-economic systems.

\FloatBarrier

\section{Discussion}\label{sec4}

The frequency and intensity of extreme events are increasing under climate change, posing escalating risks to biodiversity, human health, and food security.
Therefore, reliable projections are critical for effective climate risk assessment, mitigation, and adaptation.
While existing GCMs, such as those in CMIP, provide valuable insights, their coarse spatiotemporal resolution and systematic biases limit their ability to represent local extremes like TCs.
Recent deep learning-based downscaling methods have improved performance, yet most remain limited to univariate outputs, neglecting multivariate dependencies and temporal resolution, both of which are critical for simulating extremes.
Many approaches also fail to address distribution shifts between training and inference datasets, undermining model performance.

To address these challenges, we developed FuXi-CMIPAlign, a generative deep learning downscaling framework that integrates Flow Matching with domain adaptation, effectively mitigating input mismatches while enhancing accuracy, stability, and generalization across emission scenarios.
FuXi-CMIPAlign enables multivariate modeling, simultaneously downscaling from 70 km daily mean data to 25 km 6-hourly resolution, using MMD loss to align feature distributions between ERA5 and EC-Earth.
Evaluations over the historical period (2005–2014) and future emission scenarios (2015–2100) demonstrate that FuXi-CMIPAlign reproduces the spatiotemporal characteristics of extreme events more faithfully than both EC-Earth and alternative deep learning baselines, such as noise-based alignment and single-variable models.
It reduces global 99th-percentile MAE by 26\%, 42\%, and 33\% for high temperature, extreme precipitation, and strong wind, respectively, with the most pronounced improvements in tropical and subtropical heat extremes, localized tropical precipitation, and cyclone-prone wind regimes.
The framework also generalizes effectively to unseen scenarios and more accurately detects and classifies TC tracks. Under SSP370, the land area fraction affected by high temperature is projected to increase by up to 20\% from 2015 to 2100, while under SSP585, strong TC track density rises by 48\% in the Northwest Pacific and 60\% in the Northeast Pacific by 2100. What's more, under high-emission scenarios, extreme precipitation events over land also become increasingly pronounced, with Northwestern Australia, an area typically characterized by tropical desert climate, projected to experience up to 35\% higher rates relative to historical levels.
Additionally, the frequency and area fraction of strong wind events over land show a counterintuitive weakening trend.

Future projections reveal substantial expansion of regions exposed to heat and extreme precipitation, particularly under the high-emission scenario SSP585.
The rapid growth of heat-affected land area and intensifying extrem precipitation underscore the accelerating impacts of greenhouse gas forcing on ecosystems and socio-economic systems.
Regional analyses further show distinct spatial patterns, including increased extreme-event frequency in climate-sensitive regions such as the Middle East, northwestern Australia, and Siberia.
Interestingly, strong winds over land are projected to decrease under SSP370 and SSP585, with the rate of decline intensifying over time, most notably in Siberia, contrary to expectations based on historical trends.

Despite these advances, several limitations remain.
FuXi-CMIPAlign struggles to address cases where input data contain large systematic errors, such as persistent precipitation biases along the ITCZ in EC-Earth, likely due to complex regional dynamics and substantial errors in EC-Earth data for this region.
GPU memory constraints also necessitate daily processing units, preventing full representation of multi-day temporal continuity.
As a result, detected TC tracks can appear discontinuous, resembling the “flickering” effect similar to that observed in computer vision applications \cite{lai2018learning}, which can reduce detection accuracy.

Future research should advance in three directions.
First, optimizing training strategies for coupled multivariate systems could improve predictions of compound extremes.
Second, extending the framework to additional Earth system components, such as oceans, biosphere, and cryosphere, would enable more comprehensive projections.
incorporating more CMIP6 models and variables, such as cloud cover, solar irradiance, and 100-meter wind speed, would enhance applicability, particularly for renewable energy assessments.
Third, EC-Earth fails to detect TCs in several oceanic regions, whereas additional TCs can be identified after downscaling with FuXi-CMIPAlign. This enhancement can be viewed as generating TCs “from scratch”, a task far more challenging than refining the structure of pre-existing TCs.
While existing TCs provide continuous temporal signals, the “from scratch” generation must also maintain temporal coherence.
Otherwise, the synthesized TCs may exhibit abrupt frame-to-frame discontinuities, deteriorating the accuracy of TC detection. 

In summary, FuXi-CMIPAlign provides a robust framework for high-resolution climate analysis and impact assessment.
By more accurately reproducing historical extremes and delivering reliable scenario-based projections, it advances both methodological frontiers in climate science and strengthens the scientific foundation for climate risk assessment, mitigation, and adaptation.

\section{Method}\label{sec3}
\subsection{Data}\label{subsec4}

The ERA5 reanalysis dataset, produced by the European Centre for Medium-Range Weather Forecasts (ECMWF), provides hourly data starting from January 1940 to the present at a spatial resolution of approximately 31 km \cite{soci2024era5}.
Renowned for its extensive coverage and high accuracy, this dataset serves as the basis of this study.
We utilize ERA5 resampled to 0.25° resolution (corresponding to 720 × 1440 grid points globally) and 6-hour temporal resolution as target data.
A coarser daily-mean version with a spatial resolution of approximately 70 km (corresponding to 256 × 512 grid points globally) is used as the conditioning input.
The dataset is divided into two subsets: 1940-2000 for model training and 2005-2014 for testing.
The ERA5 variables include 2-meter temperature (T2M), 10-meter u wind component (U10M), 10-meter v wind component (V10M), mean sea level pressure (MSL), 6-hour accumulated total precipitation (TP), and geopotential (Z) at 250 hPa (Z250) and 500 hPa (Z500).
The inclusion of variables such as Z250 and Z500 is specifically aimed at detecting TC-related circulation features.

EC-Earth, a CMIP6 GCM developed collaboratively by a consortium of European meteorological and research institutions \cite{doscher2021ec}, is essential tool for understanding and predicting climate variability and climate change.
We employ its standard configuration with daily outputs at 70 km resolution (256 × 512 grid points).
Historical simulations from 1940 to 2000 are used for model training and 2005-2014 for testing.
Future projections from ScenarioMIP (2015-2100) under four shared socioeconomic pathways (SSP126, SSP245, SSP370, and SSP585) are used for inference and analysis of extreme events.
The CMIP6 variables include near-surface air temperature (TAS), eastward near-surface wind (UAS), northward near-surface wind (VAS), and sea level pressure (PSL), which correspond to ERA5 variables T2M, U10M, V10M, and MSL, respectively.
Additional variables include precipitation rate (PR), and geopotential height (ZG).
For consistency, all variables are referenced using ERA5 nomenclature throughout this paper.

We also incorporate topographic data from the Global Elevation and Continental-scale Ocean Bathymetry (GENCO), a high-resolution global digital elevation dataset integrating terrestrial and seafloor topography, commonly utilized in topographic, oceanic, and environmental science research \cite{mayer2018nippon}.
This dataset integrates land elevation and ocean bathymetry data, offering a unified reference framework for analyzing Earth's surface characteristics.
The original 30 m GENCO data are resampled to 0.25$^\circ$ resolution for this study while preserving elevation extrema.

Only month of year and day of year are used as temporal features, which are encoded into the model using sine-cosine transformations.
A summary of all datasets, variables, abbreviations, and their respective roles is presented in the table \ref{tab:meteorological_variables}.

\begin{table}[h]
\caption{Overview of data sources, variables, abbreviations, and their roles in the study. The table lists variables from the ERA5 reanalysis dataset (used as labels and conditions during training), European Consortium-Earth simulations (EC-Earth, used as conditions during testing), Global Elevation and Continental-scale Ocean Bathymetry (GENCO, used as elevation conditions), and temporal information (month and day of year). Resolution details fpr each dataset are also provided.}\label{tab:meteorological_variables}
\begin{tabular*}{\textwidth}{@{\extracolsep\fill}llll}
\toprule
Source & Full name & Abbreviation & Role \\
\midrule
ERA5 & 2-meter temperature & T2M & label (6hourly/0.25$^\circ$) \& condition (daily/70km) \\
 & 10-meter u wind component & U10M & label (6hourly/0.25$^\circ$) \& condition (daily/70km) \\
 & 10-meter v wind component & V10M & label (6hourly/0.25$^\circ$) \& condition (daily/70km) \\
 & mean sea-level pressure & MSL & label (6hourly/0.25$^\circ$) \& condition (daily/70km) \\
 & total precipitation & TP & label (6hourly/0.25$^\circ$) \& condition (daily/70km) \\
 & geopotential & Z & label (6hourly/0.25$^\circ$) \& condition (daily/70km) \\
EC-Earth & Near-Surface Air Temperature & TAS & condition (daily/70km) \\
 & Eastward Near-Surface Wind & UAS & condition (daily/70km) \\
 & Northward Near-Surface Wind & VAS & condition (daily/70km) \\
 & Sea Level Pressure & PSL & condition (daily/70km) \\
 & Precipitation Flux & PR & condition (daily/70km) \\
 & Geopotential Height & ZG & condition (daily/70km) \\
GENCO & Digital Elevation Model & DEM & condition (0.25$^\circ$) \\
Temporal & Month of year & Month & condition \\
 & Day of year & Doy & condition \\
\botrule
\end{tabular*}
\end{table}

Except for digital elevation model (DEM) data, which is normalized using the min-max method, all other variables are processed using z-score normalization. Special preprocessing is applied to TP and Z during this process to address their distinct statistical characteristics.

To mitigate the challenges in model training caused by the long-tail distribution of TP, we first apply a logarithmic transformation, $log(x+1)$, followed by normalization of the transformed data.
Additionally, the PR and ZG variables from EC-Earth are converted to align with ERA5 variables. Specifically, PR is multiplied by 86,400 (the number of seconds in a day) to obtain daily accumulated total precipitation, while ZG is multiplied by the gravitational acceleration constant (9.81 $\text{m/s}^2$) to produce Z.

\subsection{Non-Blind statistical downscaling with domain adaptation}
\label{subsec5}

Due to the distinct origins of ERA5 and EC-Earth, the degradation kernel mapping from ERA5 to EC-Earth is unknown, making a blind super-resolution task.
In computer vision, a "degradation kernel" describes the process by which high-resolution data (e.g., ERA5) is systematically transformed into lower-resolution data (such as EC-Earth) through smoothing, introducing noise, or other unknown degradation mechanisms that reduce the quality or resolution of the original data in ways that are not fully understood.
Under this setting, direct end-to-end super-resolution performs poorly, as detailed in Supplement Information section 3.2.5.
Inspired by previous studies that performed simple processing of training and inferring inputs in the spatial or frequency domain \cite{bano2022downscaling,vandal2017deepsd,hess2025fast}, we reformulate the complex blind super-resolution task into a two-step pipeline: non-blind super-resolution followed by domain adaptation.
During the training process, the model learns to reconstruct high-resolution ERA5 fields (0.25$^\circ$/6hourly) from degraded versions (70km/daily) obtained by interpolation, while simultaneously aligning the feature distributions of degraded ERA5 and EC-Earth.
The ultimate goal is to map daily 70 km EC-Earth inputs, such as T2M, U10M, V10M, MSL, TP, Z250, Z500, DEM, and temporal features, to 6 hourly outputs at 0.25$^\circ$ resolution for T2M, U10M, V10M, MSL, TP, Z250, and Z500.

\begin{figure}[h]
\centering
\includegraphics[width=0.9\textwidth]{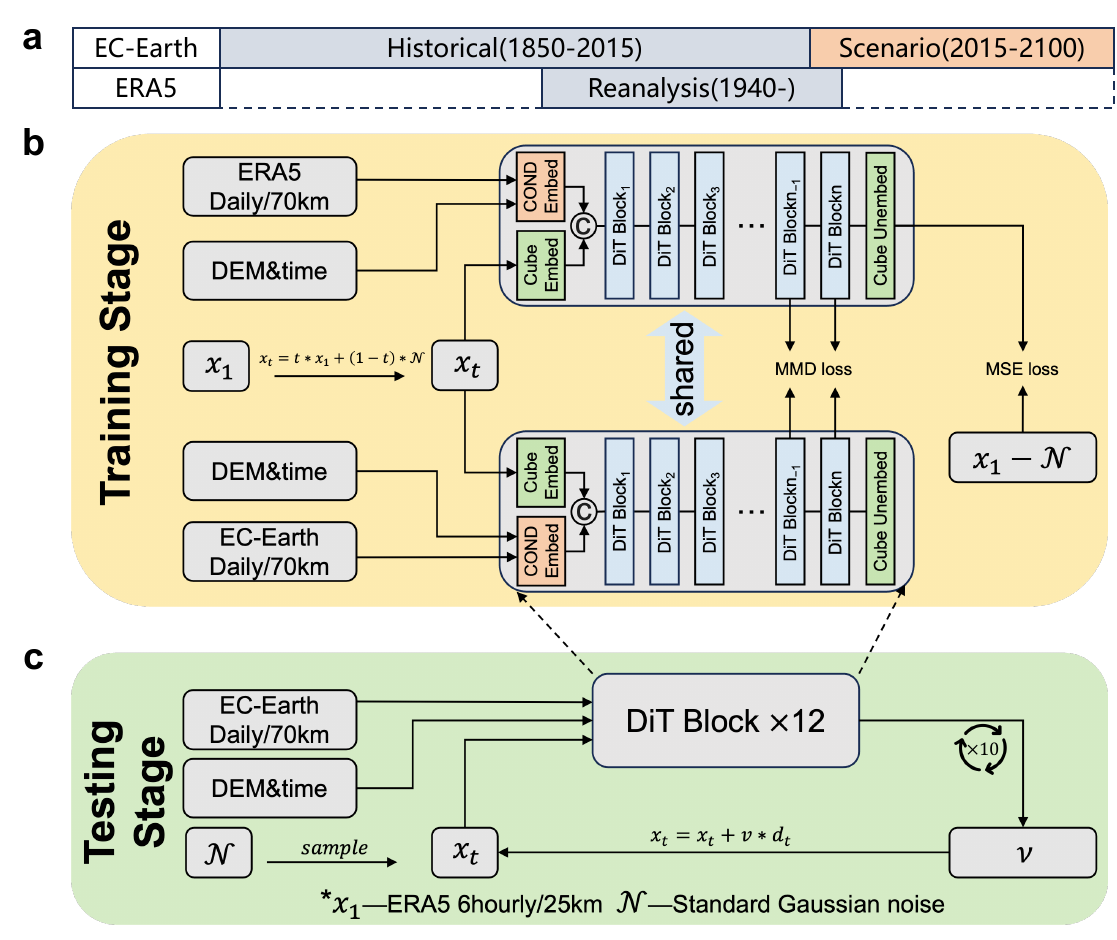}
\caption{Overview of the FuXi-CMIPAlign's architecture. \textbf{a,} Data sources and timeline, including European Consortium-Earth (EC-Earth, historical and scenario data) and ERA5. The overlapping period between the Historical and Reanalysis datasets spans from 1940 to 2014. In this study, the data from 1940 to 2000 were utilized as the training set, while the years 2001 and 2002 were designated as the validation set to monitor the training process. The period from 2005 to 2014 was employed as the test set to evaluate the performance of the model. \textbf{b,} Training stage: The model learns to reconstruct ERA5 fields at 6-hourly, 0.25$^\circ$ resolution from degraded ERA5 inputs at daily, 70km resolution, while aligning feature distributions between degraded ERA5 and EC-Earth using Maximum Mean Discrepancy (MMD) loss. Digital elevation model (DEM) and temporal features are included as additional conditions. \textbf{c,} Testing stage: The trained model downscales EC-Earth (daily, 70km resolution), using DEM and temporal features to generate high-resolution outputs (ERA5 variables at 6-hourly, 0.25$^\circ$ resolution) through iterative refinement. Gaussian noise is utilized to initialize the sampling process.}\label{FIG4}
\end{figure}

Fig. \ref{FIG4} provides an overview of the FuXi-CMIPAlign architecture.
The training stage incorporates two key components: a Flow Matching-based generative framework \cite{lipman2022flow} and a feature alignment strategy leveraging Maximum Mean Discrepancy (MMD) \cite{gretton2006kernel} (see Fig. \ref{FIG4}b).

\FloatBarrier

Flow Matching is a diffusion-based framework that learns a continuous velocity field between distributions, efficiently capturing the dynamic transformations.
Unlike conventional diffusion models (e.g., DDPM), Flow Matching directly models the velocity of distribution changes, significantly reducing the number of sampling steps while enhancing the fidelity and stability of the generated outputs.
In this study, we treat one day as the smallest sampling unit.
The target distribution consists of high-resolution ERA5 data (6-hourly, 0.25$^\circ$), and the conditional input includes low-resolution data (daily, 70 km) from two domains: degraded ERA5 (source domain) and EC-Earth (target domain).

The Flow Matching backbone is built upon a vanilla Diffusion Transformer (DiT), augmented with a condition embedding module, a cube embedding module at the head, a cube unembedding module at the tail, and 12 stacked DiT Blocks in the middle.
Specifically, the condition embedding module first interpolates the source domain conditions into a shape of $7 \times 720 \times 1440$ .
Temporal features are encoded using sine and cosine functions, expanded across spatial dimensions to $4 \times 720 \times 1440$, and concatenated with the DEM (1 channel), producing a 12-channel tensor $12 \times 720 \times 1440$ (12=7+4+1).
A two-dimensional (2D) convolutional layer (kernel size 3, stride 1, and padding 1, denoted k3s1p1 in Figure \ref{FIG4}) transforms this into a condition feature map of shape $32 \times 720 \times 1440$.
In the cube embedding module, to reduce the computational cost of three-dimensional (3D) convolutions, the noisy high spatiotemporal ERA5 data ($4 \times 7 \times 720 \times 1440$) is split along the temporal dimension into four slices.
Each slice is processed independently by a 2D convolutional layer (k3s1p1), producing target features of $32 \times 720 \times 1440$.
Finally, these are concatenated with condition features and passed through another 2D convolutional layer (k3s1p1) to reduce the number of channels to 32.
Standard Gaussian noise ($\mathcal{N}$) is added according to 
\begin{equation}
x_t = t \times x_1 + (1 - t) \times \mathcal{N}
\label{x_t_equation1}
\end{equation}
where $t \in [0,1]$, $x_1$ represents 6-hourly/0.25$^\circ$ ERA5 data, and $x_t$ represents the ERA5 data with added noise.

Before entering the DiT blocks, features undergo patch embedding (patch size = 16, hidden dimension = 768).
The backbone consists of 12 stacked DiT blocks, each with 6 attention heads.
Additionally, the noisy step is encoded using sine and cosine functions, and repeated to match the feature dimensions, and added element-wise to the features.
After being processed by the DiT blocks, features are ultimately restored to the 2D spatial domain through patch unembedding, followed by cube unembedding.
The cube unembedding module mirrors the cube embedding module: four independent 2D convolutional layers (k3s1p1) reconstruct the outputs for the four temporal slices.
The final output, velocity $v$, is compared with the target $x_1-\mathcal{N}$ using mean squared error (MSE):
\begin{equation}
\mathcal{L}_{\text{MSE}} = \|v - (x_1 - \mathcal{N})\|^2
\label{MSE_equation}
\end{equation}

For EC-Earth inputs, the same model structure and processing procedures are applied.
This ensures that two distinct hidden feature sets are extracted from the final three layers of the DiT blocks, each conditioned separately on the source and target domains. These two feature sets are projected to a common Hilbert space, where the MMD is computed and incorporated into the optimization objective.
The MMD loss is defined as:
\begin{equation}
\textrm{L}_\textrm{MMD} = \frac{1}{\textrm{N}^2} \sum_{i=1}^\textrm{N} \sum_{j=1}^\textrm{N} k(f_\text{src}^i, f_\text{src}^j) + \frac{1}{\textrm{M}^2} \sum_{i=1}^\textrm{M} \sum_{j=1}^\textrm{M} k(f_\text{tgt}^i, f_\text{tgt}^j) - \frac{2}{\textrm{N}\textrm{M}} \sum_{i=1}^\textrm{N} \sum_{j=1}^\textrm{M} k(f_\text{src}^i, f_\text{tgt}^j)
\label{MDD_loss}
\end{equation}
where $\textbf{S}_\text{src} = \{f_\text{src}^1, f_\text{src}^2, \dots, f_\text{src}^n\}$ and $\textbf{S}_\text{tgt} = \{f_\text{tgt}^1, f_\text{tgt}^2, \dots, f_\text{tgt}^m\}$ represent the source and target domain feature sets, respectively. Here, \(i\) and \(j\) are the indices for $\textbf{S}_\text{src}$ and $\textbf{S}_\text{tgt}$ .
$\textrm{N}$ and $\textrm{M}$ denote the sample sizes in each domain.
Since ERA5 and EC-Earth have the same number of samples in the training set, $\textrm{N}$ and $\textrm{M}$ can be set to \( \text{batchsize} \). And $k(f_\text{src}, f_\text{tgt})$ is the Gaussian kernel function defined as
\begin{equation}
k(f_\text{src}, f_\text{tgt}) = \exp\left(-\frac{\|f_\text{src} - f_\text{tgt}\|^2}{2\sigma^2}\right)
\label{bandwidth}
\end{equation}
where $\sigma$ is the bandwidth.

Thus, the overall loss function combines the MSE and MMD:
\begin{equation}
\textrm{L} = \textrm{L}_\textrm{MSE} + \textrm{L}_\textrm{MMD}
\label{loss}
\end{equation}

During the testing stage, the model takes as input the initial noise field, EC-Earth, DEM, and temporal features.
The initial high-resolution noise $x_t$ is sampled from a standard Gaussian distribution.
The iterative denoising process follows:
\begin{equation}
x_t' = x_t + v \times d_t
\label{x_t_equation}
\end{equation}
where $x_t$ and $x_t'$ represent the downscaled fields before and after the update, $v$ denotes the velocity field predicted by the model, and $d_t$ is the time differential, defined as the reciprocal of the denoising step count.

The model was trained on two NVIDIA Tesla A100 GPUs.
Training employed the AdamW optimizer with an initial learning rate of $1\times10^{-4}$, decayed by 10\% every 50 epochs using a StepLR scheduler, over a total of 200 epochs.
During testing, the number of denoising steps was set to 10.


\subsection{Extreme detection}\label{subsec6}

Following \cite{karl1999clivar}, we defined extreme events (extreme temperature, precipitation, and wind) using the percentiles of the 1979–2000 baseline distribution.

The 99th-percentile of ERA5 data for the reference period 1979-2000 was first calculated for each grid point.
Values exceeding this threshold were assigned a value of 1, and all others 0.
Summing across all grid points at a each time step yields the total number of extreme occurrences, while dividing by the total number of grid points provided the area fraction of the spatial domain affected by extreme events.

TC tracks were detected using TempestExtremes \cite{ullrich2021tempestextremes}, a versatile toolkit developed by the Climate and Global Change Research Team at the University of California, Davis.
TempestExtremes is designed for detecting and tracking extreme weather events in large-scale climate datasets.
We applied TempestExtremes to both daily and 6-hourly ERA5 data, and observed substantial differences in the number of detected TCs.
On average, the daily data captured only 4 TCs per year, while the 6-hourly data identified approximately 58 more TCs annually. (see Supplementary Information Figure 1), underscoring the critical role of temporal resolution in accurately detecting tropical cyclones.
Parameters settings for the detection and suture operators are detailed in Supplementary Information Table 1.
TC intensity was defined following the national standard \cite{GBT19201_2006}, which classifies events based on the maximum WS10M along each track.
The TC intensity categories used in this study are summarized in Table \ref{TC_intensity_scale}:

\begin{table}[h!]
    \centering
    \caption{Tropical cyclone (TC) intensity categories are defined based on the national standard \cite{GBT19201_2006}, which classifies TCs based on their maximum sustained wind speed over the entire lifecycle.}
    \begin{tabular}{ll}
        \toprule
        \textbf{Maximum WS10M (m/s)} & \textbf{Grade} \\
        \midrule
        10.8--17.1 & Tropical Depression \\
        17.2--24.4 & Tropical Storm \\
        24.5--32.6 & Severe Tropical Storm \\
        32.7--41.4 & Typhoon (Hurricane) \\
        41.5--50.9 & Severe Typhoon (Hurricane) \\
        \bottomrule
    \end{tabular}
    \label{TC_intensity_scale}
\end{table}

\subsection{Evaluation method}\label{subsec7}


The evaluation framework is centered on quantile analysis.
Two sets of spatial quantiles derived from ERA5 serve as reference for the evaluation.
The first set, calculated from ERA5 data for 2005-2014, is used to assess the model performance during this historical period.
The second set, derived from ERA5 data from 1979-2000, serves as a historical benchmark for assessing extreme events projected by FuXi-CMIPAlign under future emission scenarios (2015-2100).

The first set of quantiles, based on the 90th, 95th, and 99th percentiles, serves as the reference. For each model, we also calculate the 90th, 95th, and 99th percentiles over the same period (2005–2014). The mean absolute error (MAE) between the model-derived quantiles and the reference quantiles is then used to evaluate the model's ability to reproduce extreme climate states during this validation period.
The MAE for a given quantile $q$ is computed as:
\begin{equation}
\text{MAE}_{q} = \frac{1}{\textrm{N}\_\textrm{i} \textrm{N}\_\textrm{j}} \sum_{i=1}^{\textrm{N}\_\textrm{i}} \sum_{j=1}^{\textrm{N}\_\textrm{j}} \left| \left( \textrm{P}^{q}_{\text{model},i,j} - \textrm{P}^{q}_{\text{ERA5},i,j} \right) \cdot w_j \right|
\label{MAE_q}
\end{equation}
where $\textrm{N}\_\textrm{i}$ and $\textrm{N}\_\textrm{j}$ represent the number of grid points in the longitude and latitude directions respectively. Since our experiment covers the entire globe, $\textrm{N}\_\textrm{i}$ is set to 1440 and $\textrm{N}\_\textrm{j}$ is set to 720. \(i\) and \(j\) are the indices for longitude and latitude directions respectively. \(\textrm{P}^{q}_{\text{model},i,j}\) and \(\textrm{P}^{q}_{\text{ERA5},i,j}\) represent the model and ERA5 quantile values at the grid point \(i,j\). \(w_j\) represent the weight of latitude.

The second set only has 99th percentile as a historical benchmark for identifying extreme grid points and analyzing spatiotemporal changes in future extremes.
The frequency of extreme events is obtained by counting their occurrences, while the affected area fraction is defined as the ratio of extreme pixels to the total number of pixels globally.


\backmatter

\section*{Data Availability}
The EC-Earth data are accessible through the Earth System Grid Federation at \url{https://esgf-ui.ceda.ac.uk/cog/search/cmip6-ceda/}.
The ERA5 reanalysis data are accessible through the Copernicus Climate Data Store at \url{https://cds.climate.copernicus.eu/}.

\section*{Code Availability}
The TempestExtremes tracker used in this study is available from \url{https://github.com/ClimateGlobalChange/tempestextremes}.
The model for FuXi-CMIPAlign used in this study is available at \url{https://zenodo.org/records/17311569} \cite{fuxi_cmipalign_2025}.

\section*{Acknowledgements}
We gratefully acknowledge the ECMWF for their efforts in producing, maintaining, and distributing the ERA5 reanalysis data.
We thank the European consortium of national meteorological services and research institutes for developing and distributing EC-Earth data.

\section*{Competing interests}
The authors declare no competing interests.











\noindent







\bibliography{sn-bibliography}

\end{document}